\documentclass[showpacs,prd,preprintnumbers,nofootinbib]{revtex4}%
\usepackage{amssymb}
\usepackage{amsmath}
\usepackage{graphicx}
\usepackage{hyperref}
\usepackage{color}
\usepackage{amsfonts}%
\providecommand{\U}[1]{\protect\rule{.1in}{.1in}}
\begin{document}

\title{Skyrmion vibrational energies with a generalized mass term}
\author{Merlin C. Davies and Luc Marleau}
\email{lmarleau@phy.ulaval.ca}
\affiliation{D\'epartement de Physique, de G\'enie Physique et d'Optique, Universit\'e
Laval, Qu\'ebec, Qu\'ebec, Canada G1K 7P4}
\date{\today}

\begin{abstract}
We study various properties of a one parameter mass term for the Skyrme model, originating from the
works of Kopeliovich, Piette and  Zakrzewski \cite{KPZ},
through the use of axially symmetric solutions obtained numerically by
simulated-annealing. These solutions allow us to observe asymptotic behaviors
of the $B=2$ binding energies that differ to those previously obtained
\cite{PZ}. We also decipher the characteristics of three distinct vibrational
modes that appear as eigenstates of the vibrational Hamiltonian. This analysis
further examine the assertion that the one parameter mass term offers a
better account of baryonic matter than the traditional mass term.

\end{abstract}

\maketitle

\section{Introduction}

Being one of the primary candidates for an effective low-energy theory of QCD,
the Skyrme model \cite{Skyrme1,Skyrme2,Skyrme3,Skyrme4} has relished from a
large amount of study after which it was realized that it possessed the same
symmetry properties as QCD in the limit of large $N_{c}$ \cite{Witten1}. The
topological solitons that appear as solutions to the model's equations of
motion are identified as mathematical representations of nuclear matter. These
solitons, or Skyrmions, are then quantized to obtain physical properties of
nuclei. Studies have shown that for the nucleon $B=1$, these calculated
properties are within a 30\% margin of error from experimental data
\cite{Adkins1}.

However, when one raises the baryonic number to any $B \geq2$, the solutions
that have been obtained thus far do not correctly describe the presumed
geometric properties of nuclear matter. Experimental data indicates that the
nucleons seem to preserve their individuality within nuclei. For example, the
$B=2$ solutions have the geometrical shape of a toroid, whereas the deuteron,
the only stable nuclei with $B=2$, has the presumed shape of two deformed nucleons
lightly bound together. Another problem is that the binding energy of the
toroidal Skyrmion is much too high, roughly $\sim80$ MeV (Recall that the
deuteron has a binding energy of 2.224 MeV), which is also a likely
contributing factor to the odd toroidal shape it possesses. Even worst, both
the geometrical shape and binding energy problems persist and are amplified as
we increase the value of $B$. For these reasons, the solitonic field
configurations within the standard Skyrme model can be viewed as too
\emph{malleable} in the sense that Skyrmions deform noticeably as they form a bound state. For the purpose of comparison here we shall use vibrational
energies as a quantifiable measure of rigidity (as opposed to malleability) of
the field and analyze the ratio $\frac{\omega_{i}}{E^{tot}}$, where
$i=br,2,3_{}$  labels the vibrational energy of the breathing mode and the two other eigenmodes, and where
$E^{tot}$ is the total energy of the soliton (more on this in Sec. III).
Reaching for higher rigidity solutions can also be understood intuitively by the
fact that every nucleon within a nuclei of $B=2$ or higher should presumably
be deformed while maintaining its  individuality within the nucleus, and not meld with others to form complex
geometrical objects. A pragmatic goal towards improving the model would be to
find extensions to the current model that point toward a solution to these
issues. Therefore, in this optic, one may consider various generalizations of
the original Skyrme Lagrangian to find which types lowers the binding
energies. In this paper, we limit our analysis to a one-parameter
generalization to the standard mass term of the Skyrme model.

In the following, we begin by reviewing the standard Skyrme model as well as
reveal the studied generalized mass term that introduces a dimensionless
parameter labeled $D$. Initially considered and analyzed by Piette and  Zakrewski \cite{PZ}, this new mass term is studied further in the context
of rigidity here using the simulated-annealing numerical algorithm to
accurately and effectively minimize the pion field configuration. These exact
solutions allow us to deepen our understanding of the dependency between
binding energies and the parameter $D$, and will also enable us to conclude
that solitonic solutions obtained through the rational map ansatz \cite{PZ}
are not good indicators of this dependence (Sec. IV). We consider solutions  for $B=1$ and
$B=2$ with axially symmetric configurations. For $B=1$, axial symmetry is an exact symmetry both for
the static solution and the
rotationally deformed solution.
Both calculation will be performed  providing a quantitative  insight on  such deformations otherwise
shown to be significant  \cite{Battye:2005,Magee,Marleau2}.  For $B=2,  $ the situation is somewhat
different. Rotational deformation breaks axial symmetry and would in principle require a full 3D computation. So only the static energy will be minimized for the \(B=2\) case  even though axially symmetric solutions were found
to be a  good approximation in \cite{Marleau2}. This was done in order to
make the numerical efforts more tractable and time efficient. Furthermore, the
same solutions are used to compute how the vibrational energies of three
eigenstates behave as we increase $D$. The methods used for this analysis is
outlined in Sec. III, and the results indicate quantitatively that the mass
term increases the rigidity of Skyrmions with $B=1$ while its effect on the binding energy depend on
(iso-)spin.

\section{The Skyrme model with an extended mass term}

The model initially proposed by Skyrme  comprised of the two
term Lagrangian
\begin{equation}
L=\int d^{3}x\left[  -\frac{F_{\pi}^{2}}{16}\,Tr\left(  L_{\mu}L^{\mu}\right)
+\frac{1}{32e^{2}}\,Tr\left(  [L_{\mu},L_{\nu}]^{2}\right)  \right]  ,
\label{eq-S1}%
\end{equation}
where $L_{\mu}=U^{\dag}\partial_{\mu}U$ are the chiral currents associated to
the three component pion fields $\mathbf{\pi}$ such that the $SU(2)$ matrix
$U$ is defined by
\begin{equation}
U=\sigma+i\mathbf{\tau}\cdot\mathbf{\pi}. \label{eq-U}%
\end{equation}
$F_{\pi}$ and $e$ are respectively the pion decay constant and the
dimensionless Skyrme parameter. The $\mathbf{\tau}$'s found in (\ref{eq-U}) are
simply the Pauli matrices, and the field $\sigma$ is an additional scalar
field that must satisfy the constraint $\sigma^{2}+\mathbf{\pi}\cdot
\mathbf{\pi}=1$ in order to avoid adding unphysical degrees of freedom and to
enable the possibility of having solitonic solutions.

Each solution to (\ref{eq-S1}) with boundary condition
\begin{equation}
U(\mathbf{r},t)\rightarrow1\quad\text{as}\quad|\mathbf{r}|\rightarrow
\infty\label{boundary}%
\end{equation}
fall into distinct topological sectors that are distinguished by their given
topological charge, or baryon number. For $B=1$, spherical symmetry can be
utilized to extract a convincing picture of a nucleon through the use of the
ansatz
\[
U=e^{i\tau_{i}\hat{x}_{i}F(r)},
\]
where one only has to solve an ordinary differential equation involving the
chiral angle $F(r)$ that (\ref{eq-S1}) brings about \cite{Adkins1}.

However, the physics that (\ref{eq-S1}) describes lacks in many respects.
First, it is possible to add as many higher order terms of the form
$Tr\,([L,L]^{n})$ as one wishes, to account for all possible interactions, and
it is simply unknown how the sum of such terms would affect the resulting
solutions \cite{Marleau1} although it is generally assumed that higher order
terms in derivatives could be neglected in the low-energy limit. One may also wonder if the two terms of (\ref{eq-S1}), among all the possible terms, are
really the two optimal ones for portraying the physics of nuclear matter.
Second, all solutions for $B\geq2$ obtained thus far do not correctly
characterize the presumed geometric properties of nuclei that are found in
nature. Third, it does not take into
account the mass the of pions, but fortunately, this problem is easier to
resolve. One can simply add a chiral symmetry breaking term proportional to
the pion mass squared, which has first been successfully introduced by 
Adkins and Nappi \cite{Adkins2}, in the form
\begin{equation}
\frac{m_{\pi}^{2}F_{\pi}^{2}}{8}\,Tr\,(\mathbf{1}-U), \label{eq-M1}%
\end{equation}
where $m_{\pi}$ is the pion mass (We assume that the three pions $\pi^{0}$,
$\pi^{+}$, and $\pi^{-}$ are of equal masses). This term has had the added
benefit of eliminating shell-like configurations and favoring energy densities
that are higher at their centers, which is more appealing since it is known
that nucleons have roughly even matter densities within their shell radii.
Hence, the Lagrangian (\ref{eq-S1}) together with the mass term (\ref{eq-M1})
is what we consider to be the standard Skyrme model. Of course, as we have
mentioned, many additions and modifications can be made, particularly to the
mass term (\ref{eq-M1}).

If we only constrain ourselves to mass terms that obey the boundary condition
(\ref{boundary}),  Kopeliovich,  Piette and  Zakrzewski
\cite{KPZ} have shown that a generalized mass term can have the form
\begin{equation}
\frac{m_{\pi}^{2}F_{\pi}^{2}}{8K}Tr\left(  1-\int_{-\infty}^{+\infty}%
g(p)U^{p}dp\right)  , \label{MassG}%
\end{equation}
where the function $g(p)$ and constant $K$ must obey
\begin{equation}
\int_{\infty}^{\infty}g(p)dp=1\quad\text{and}\quad K=\int_{\infty}^{\infty
}g(p)p^{2}dp.\label{MassG2}%
\end{equation}
However, of the many mass terms that are evidently possible, we will study a
particular one-parameter family that has the additional property of
disfavoring shell-like configurations as did the standard mass term
(\ref{eq-M1}). One then hope that this \emph{new} term might also improve the
overall properties of Skyrmions, such as providing a better account of their
binding energies. This one-parameter mass term is
\begin{equation}
\frac{m_{\pi}^{2}F_{\pi}^{2}}{8(1-5D)}\,Tr\,(\mathbf{1}-U-D(U^{2}-U^{3})),
\label{eq-MD}%
\end{equation}
is based on the function 
\begin{equation*}
g(p)=\delta (p-1)+D\,(\delta (p-2)-\delta (p-3))
\end{equation*}
where \(K=1-5D\) and the parameter $D$ can span the range $\left[  \,0,0.2\,\right[  $
\vspace{1pt} . At $D=0.2$, we note that the mass term is infinite.

Following their original proposal of the general term (\ref{MassG}),  Piette
and  Zakrzewski \cite{PZ} studied (\ref{eq-MD}) with the use of rational
map (RM) solutions. In Sec. IV, we will show the results of a similar study
that use exact numerical solutions instead of approximate RM map solutions.
This will demonstrate the limitations of the rational map ansatz when it comes
to give a precise measurement of the binding energies as a function of $D$,
especially as it approaches the critical value of 0.2.

Before we discuss binding energies, we will describe in the following section
how the study of vibrational modes, a worthy study in its own right, can lead
us to quantitatively understand how the $B=1$ solutions gradually get more
rigid as $D$ increases. The methods used in this paper for calculating the
vibrational energies were previously outlined by Hadjuk andt Schwesinger \cite{Haj1},
which can be contrasted with the techniques of Barnes et al. \cite{Barnes1}
and of Lin and Piette \cite{LP} where the latter performed their vibrational
calculations using rational map solutions.

\section{Vibrational modes and their energies}

Originally put forward by Hajduk and Schwesinger \cite{Haj1}, the method for obtaining
the vibrational modes and their energies begins by performing the global scale
transformation
\begin{equation}
\sigma(x) \rightarrow\sigma(\beta_{k} x_{k}) \quad\text{and} \quad\pi_{a}(x)
\rightarrow\pi_{a} (\beta_{k} x_{k}) , \label{scaling}%
\end{equation}
where the $\beta_{k}$ are scaling parameters, and where we have assumed that
the scaling is uniform with respect to the Cartesian coordinates $x$, $y$ and
$z$. The advantage of such a scaling lies in its simplicity both in its
mathematical and numerical treatment. Its disadvantage is that it is certainly
not as general as would be an arbitrary local scale transformation. However,
it allows to compute vibrational modes that probe the global \emph{rigidity}
of Skyrmions, giving us a clearer indication whether or not (and how) the
field configurations are getting more \emph{rigid} with respect to the
parameter $D$. Substituting (\ref{scaling}) into (\ref{eq-S1}) together with
the new mass term (\ref{eq-MD}) yields a Lagrangian of the form
\begin{equation}
L = \frac{1}{2} M_{ij}(\beta) \frac{\dot{\beta_{i}} \dot{\beta_{j}}}{\beta_{i}
\beta_{j}} - V(\beta),
\end{equation}
where the matrices $M_{ij}(\beta)$ and $V(\beta)$ are obtained by direct
inspection after all substitutions are made. Since we are only concerned with
small amplitude oscillations, we perform an expansion around the minimum of
$V$ by taking
\begin{align}
\beta_{i}  &  = \beta^{0}_{i} e^{\eta_{i}} = \beta^{0}_{i} (1 + \eta_{i} +
\frac{1}{2} \eta^{2}_{i} +...)\\
\dot{\beta_{i}}  &  = \beta^{0}_{i} \dot{\eta_{i}},
\end{align}
which results in the expansion of $M_{ij}$ and $V$ as
\begin{align}
M_{ij}(\beta)  &  = M_{ij} \Big|_{\beta^{0}_{1}, \, \beta^{0}_{2}, \,
\beta^{0}_{3}} + \left(  \eta_{k} \partial_{k} M_{ij} \right)  \Big|_{\beta
^{0}_{k}} +...\nonumber\\
&  \approx M^{0}_{ij} + \left(  \eta_{k} \partial_{k} M_{ij} \right)
\Big|_{\beta^{0}_{k}}\\
V(\beta)  &  = V \Big|_{\beta^{0}_{1}, \, \beta^{0}_{2}, \, \beta^{0}_{3}} +
\frac{1}{2} \eta_{i} \eta_{j} (\partial_{i} \partial_{j} V ) \Big|_{\beta
^{0}_{1}, \, \beta^{0}_{2}, \, \beta^{0}_{3}} +...\nonumber\\
&  \approx V^{0} + \frac{1}{2} \eta_{i} \eta_{j} v_{ij}.
\end{align}
Keeping only terms up to order $O(\eta^{2})$ gives us the Lagrangian
\begin{equation}
L = \frac{1}{2} M^{0}_{ij}(\beta) \dot{\eta_{i}} \dot{\eta_{j}} - \frac{1}{2}
\eta_{i} \eta_{j} v_{ij}. \label{eq-L}%
\end{equation}
In order to compute the vibrational Hamiltonian we must now find a coordinate
transformation that satisfies
\begin{equation}
A^{T} M^{0} A = 1,
\end{equation}
where $\eta= A \xi$. When such a transformation is found, it gives
\begin{equation}
L = \frac{1}{2} \dot{\xi_{i}}\dot{\xi_{j}} - \frac{1}{2} (A^{T} v A)_{ij}
\xi_{i} \xi_{j},
\end{equation}
which can simply be turned into the Hamiltonian
\begin{equation}
H_{vib} = \frac{1}{2} \sum_{i} \frac{\partial^{2}}{\partial\xi^{2}_{i}} +
\frac{1}{2} \left(  A^{T} v A \right)  _{ij} \xi_{i} \xi_{j}.
\end{equation}
We now need to diagonalize the matrix $(A^{T} v A )_{ij}$ in order to obtain
the vibrational eigenstates and eigenvalues. Thus, we need to solve the
eigenvalue equation
\begin{equation}
\left(  B^{T} A^{T} v AB \right)  _{ij} = \omega^{2}_{i} \delta_{ij},
\label{diag}%
\end{equation}
where the matrix $B$ must satisfy $B^{T} B = 1$. The energies associated to
the eigenstates $B_{1j}$, $B_{2j}$, et $B_{3j}$ are then
\begin{equation}
E_{i}^{vib} = \left(  n_{i} + \frac{1}{2} \right)  h \omega_{i}.
\label{Ew}\end{equation}
However, in our study, we set the zero-point energy to zero, in other words we
set
\begin{align}
E_{i}^{vib} = h n_{i} \omega_{i}%
\end{align}
because the zero-point energy of vibrational modes is ill-defined. This is
procedurally how we obtained our eigenstates (eigen-modes) and eigenvalues
(eigen-energies) from our numerical solutions. But, before we discuss the
results of our vibrational analysis of $B=1$ solitons, we must briefly
describe what exactly are the numerical solutions we are working with.
For simplicity, we shall from hereon identify vibrational frequencies \( \omega_{i}\) as vibrational energies although they are not exactly the same (see eqn. (\ref{Ew})).

Beginning from the Lagrangian (\ref{eq-S1}), we added the mass term
(\ref{eq-MD}), yielding the \emph{static} energy
\begin{align}
E_{s}^{B}  &  =\int d^{3}x\left[  -\frac{F_{\pi}^{2}}{16}\,Tr\left(
L_{i}L^{i}\right)  +\frac{1}{32e^{2}}\,Tr\left(  [L_{i},L_{j}]^{2}\right)
\right] \nonumber\\
&  +\frac{m_{\pi}^{2}F_{\pi}^{2}}{8(1-5D)}\,Tr\,(\mathbf{1}-U-D(U^{2}-U^{3})),
\label{EStatic}%
\end{align}
where $i$ and $j$ run over spatial components only and $B$ is the baryonic
number. The minimal energy Skyrmion for $B=1$ and $B=2$ turns out to have
spherical and axial symmetry respectively. Since we are only interested by
these values of $B$, the general solution will be cast in the form of the
axial ansatz
\begin{equation}
\sigma=\psi_{3}\hspace{1cm}\pi_{1}=\psi_{1}\cos n\theta\hspace{1cm}\pi
_{2}=\psi_{1}\sin n\theta\hspace{1cm}\pi_{3}=\psi_{2} \label{eq:solaxiale}%
\end{equation}
introduced in \cite{Krusch:2004uf} where $\boldsymbol{\psi}(\rho,z)=(\psi
_{1},\psi_{2},\psi_{3})$ is a three-component unit vector that is independent
of $\theta$. The boundary conditions at infinity implies that
$\boldsymbol{\psi}\rightarrow\left(  0,0,1\right)  $ as $\rho^{2}%
+z^{2}\rightarrow\infty$. Moreover, we must impose that $\psi_{1}=0$ and
$\partial_{\rho}\psi_{2}=\partial_{\rho}\psi_{3}=0$ at $\rho=0$.

With the axial ansatz (\ref{eq:solaxiale}), we can also set the scaling length
and energy in units of ${2\sqrt{2}}/{eF_{\pi}}$ and ${F_{\pi}}/{2\sqrt{2}}$
respectively\footnote{We have used ${2\sqrt{2}}/{eF_{\pi}}$ and ${F_{\pi}%
}/{2\sqrt{2}}$ as units of length and energy respectively.}, giving
expressions for the static energy and the baryon number that read
\begin{align}
E_{s}^{B}  &  =-\int\text{d}^{3}x\mathcal{L}_{S}\nonumber\\
&  =2\pi\left(  \frac{F_{\pi}}{2\sqrt{2}e}\right)  \int\text{d}z\text{d}%
\rho\rho\biggl\{\left(  \partial_{\rho}\boldsymbol{\psi}\cdot\partial_{\rho
}\boldsymbol{\psi}+\partial_{z}\boldsymbol{\psi}\cdot\partial_{z}%
\boldsymbol{\psi}\right)  \left(  1+n^{2}\frac{\psi_{1}^{2}}{2\rho^{2}}\right)
\nonumber\\
&  +\frac{1}{2}|\partial_{z}\boldsymbol{\psi}\times\partial_{\rho
}\boldsymbol{\psi}|^{2}+n^{2}\frac{\psi_{1}^{2}}{\rho^{2}}+\frac{2\beta^{2}%
}{(1-5D)}\left(  1-\psi_{3} \right)  \left(  1 +D\left(  1 - 2\psi_{3}%
-4\psi_{3}^{2}\right)  \right)  \biggr\} \label{eq:Estataxiale}%
\end{align}

\begin{equation}
B=\frac{n}{\pi}\int\text{d}z\text{d}\rho\,\psi_{1}|\partial_{\rho
}\boldsymbol{\psi}\times\partial_{z}\boldsymbol{\psi}| \label{eq:Bcyl}%
\end{equation}
with $\beta=\frac{2\sqrt{2}m_{\pi}}{eF_{\pi}}$.

Also, for all our minimizations, we fixed the constants $F_{\pi}$, $e$, and
$m_{\pi}$ to
\[
F_{\pi}=129\,\text{MeV},\quad e=5.44\,\text{MeV}^{-1},\quad\text{and}\quad
m_{\pi}=138\,\text{MeV}.
\]
The values of $F_{\pi}$ and $e$ were set according to ref. \cite{KPZ} for
comparative purposes. At such values, the Skyrme model reproduces the mass of
the nucleon and of the delta when one discounts the mass term \cite{Adkins1}.
Regarding the mass term, $m_{\pi}$ is set to its experimental value and the
parameter $D$ varies in order to probe how the mass term in (\ref{EStatic})
might affect the Skyrmion's properties. Of course, when the mass term is
switched on, the prediction for the nucleon and delta masses will deviate from
their experimental values. This could be corrected with an appropriate choice
of $F_{\pi}$ and $e$, however here we shall retain $F_{\pi}=129$ MeV and
$e=5.44$ MeV$^{-1}$, since we are more interested in certain ratios of
energies than in the actual values of energies, and comparisons with earlier
works are easier.

Physical states such as the nucleon and the deuteron require an additional
contribution to the energy, the rotational and isorotational energy due to
spin and isospin of these states. For the nucleon, this is done by fixing the
quantum numbers of spin and isospin to $I=J=\frac{1}{2}$, and assuming axial
symmetry, it leads to the total isorotational and rotational energy of the
form
\begin{equation}
E_{rot}^{1}=\frac{1}{4}\left[  \frac{\left(  1-\frac{W_{11}}{U_{11}}\right)
^{2}}{V_{11}-\frac{W_{11}^{2}}{U_{11}}}+\frac{1}{U_{11}}+\frac{1}{2U_{33}%
}\right]  . \label{ERot}%
\end{equation}
Similarly for the deuteron one gets%
\begin{equation}
E_{rot}^{2}=\frac{1}{V_{11}}. \label{ERot2}%
\end{equation}
Here $U_{11}$, $U_{33}$, $V_{11}$, and $W_{11}$ are moments of inertia which
follows the definition in the works of Houghton and Magee \cite{Magee} and
Fortier and Marleau \cite{Marleau2}. Accordingly the components of these
inertia tensors are
\begin{align}
U_{11}  &  =2\pi\left(  \frac{2\sqrt{2}}{e^{3}F_{\pi}}\right)  \int
\text{d}z\text{d}\rho\rho\biggl\{\psi_{1}^{2}+2\psi_{2}^{2}+\frac{1}%
{2}\biggl[\left(  \partial_{\rho}\boldsymbol{\psi}\cdot\partial_{\rho
}\boldsymbol{\psi}+\partial_{z}\boldsymbol{\psi}\cdot\partial_{z}%
\boldsymbol{\psi}+n^{2}\frac{\psi_{1}^{2}}{\rho^{2}}\right)  \psi_{2}^{2}\\
&  +\left(  \partial_{\rho}\psi_{3}\right)  ^{2}+\left(  \partial_{z}\psi
_{3}\right)  ^{2}+n^{2}\frac{\psi_{1}^{4}}{\rho^{2}}%
\biggr]\biggr\},\label{eq:U11cyl}\\
U_{33}  &  =2\pi\left(  \frac{2\sqrt{2}}{e^{3}F_{\pi}}\right)  \int
\text{d}z\text{d}\rho\rho\psi_{1}^{2}\left(  \partial_{\rho}\boldsymbol{\psi
}\cdot\partial_{\rho}\boldsymbol{\psi}+\partial_{z}\boldsymbol{\psi}%
\cdot\partial_{z}\boldsymbol{\psi}+2\right)  ,\label{eq:U33cyl}\\
V_{11}  &  =2\pi\left(  \frac{2\sqrt{2}}{e^{3}F_{\pi}}\right)  \int
\text{d}z\text{d}\rho\rho\biggl\{|\rho\partial_{z}\boldsymbol{\psi}%
-z\partial_{\rho}\boldsymbol{\psi}|^{2}\left(  1+n^{2}\frac{\psi_{1}^{2}%
}{2\rho^{2}}\right)  +z^{2}n^{2}\frac{\psi_{1}^{2}}{\rho^{2}}+\frac{1}%
{2}\left(  \rho^{2}+z^{2}\right)  |\partial_{\rho}\boldsymbol{\psi}%
\times\partial_{z}\boldsymbol{\psi}|^{2}\biggr\},\label{eq:V11cyl}\\
W_{11}  &  =2\pi\left(  \frac{2\sqrt{2}}{e^{3}F_{\pi}}\right)  \int
\text{d}z\text{d}\rho\rho\biggl\{\left[  \psi_{1}\left(  \rho\partial_{z}%
\psi_{2}-z\partial_{\rho}\psi_{2}\right)  -\psi_{2}\left(  \rho\partial
_{z}\psi_{1}-z\partial_{\rho}\psi_{1}\right)  \right]  \biggl(1+\frac{1}%
{2}\left[  \left(  \partial_{z}\psi_{3}\right)  ^{2}+\left(  \partial_{\rho
}\psi_{3}\right)  ^{2}+\frac{\psi_{1}^{2}}{\rho^{2}}\right]  \biggr)\\
&
\phantom{=2\pi\left(\frac{2\sqrt{2}}{e^3F_\pi}\right)\int \text{d}z\text{d}\rho\rho\biggl\{}+\frac
{\psi_{3}}{2}\left(  z\partial_{z}\psi_{3}+\rho\partial_{\rho}\psi_{3}\right)
\left[  \partial_{\rho}\psi_{2}\partial_{z}\psi_{1}-\partial_{\rho}\psi
_{1}\partial_{z}\psi_{2}\right]  +\frac{z\psi_{1}\psi_{2}}{2\rho}\left(
2+\partial_{\rho}\boldsymbol{\psi}\cdot\partial_{\rho}\boldsymbol{\psi
}+\partial_{z}\boldsymbol{\psi}\cdot\partial_{z}\boldsymbol{\psi}\right)
\biggr\}. \label{eq:W11cyl}%
\end{align}
Finally, the nucleon and deuteron mass are the sum of the appropriate static
and rotational energies $E_{B}^{s}+E_{B}^{rot}$ with \(B=1,2\) respectively.

\begin{figure}[ptbh]
\centering
\begin{tabular}{|c|c|}\hline
\includegraphics[width=0.2\textwidth]{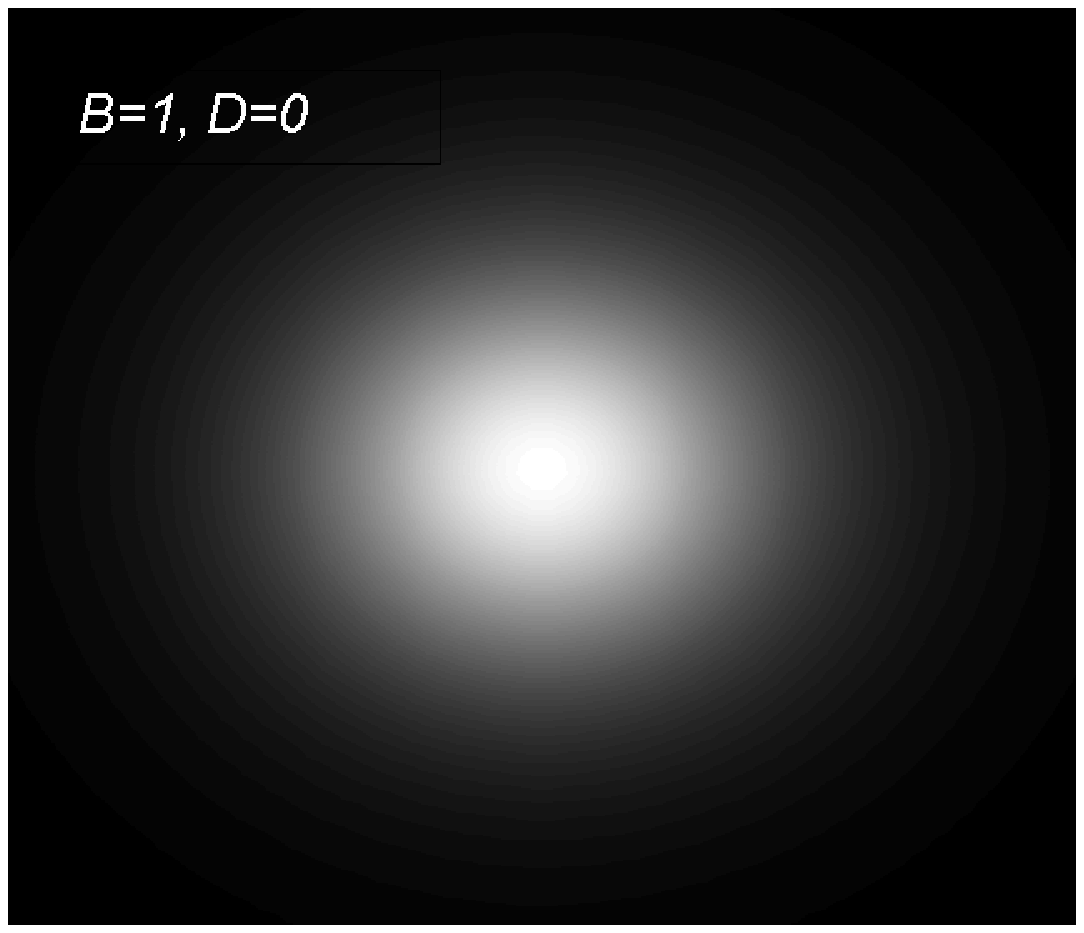} & 
\includegraphics[width=0.2\textwidth]{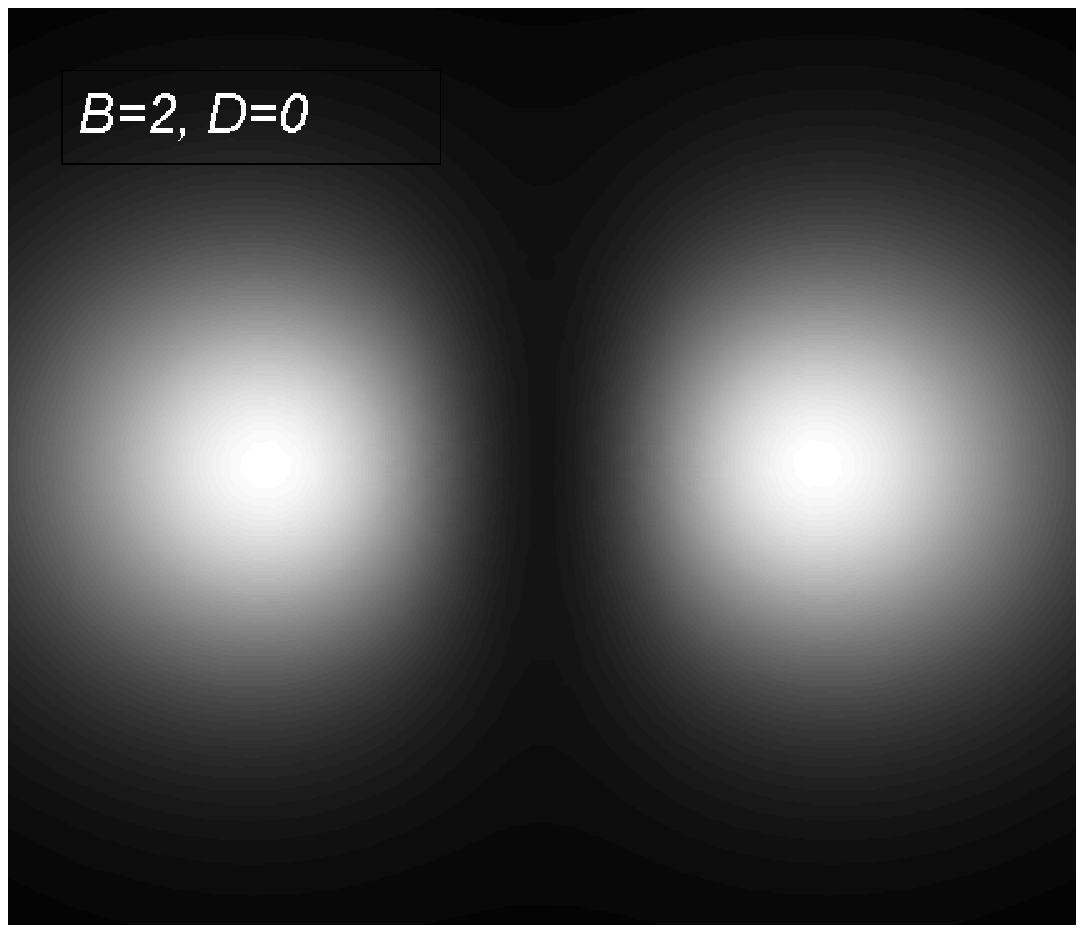} \\\hline
\includegraphics[width=0.2\textwidth]{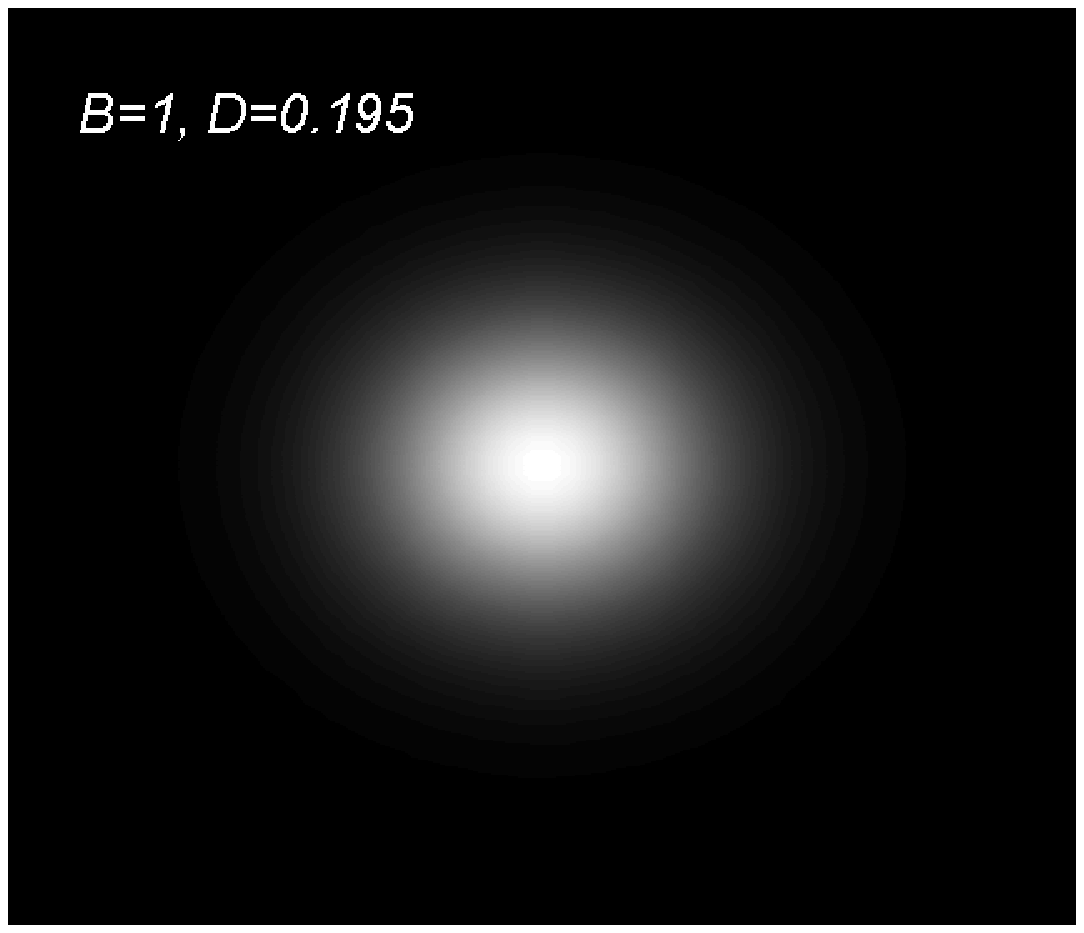} & 
\includegraphics[width=0.2\textwidth]{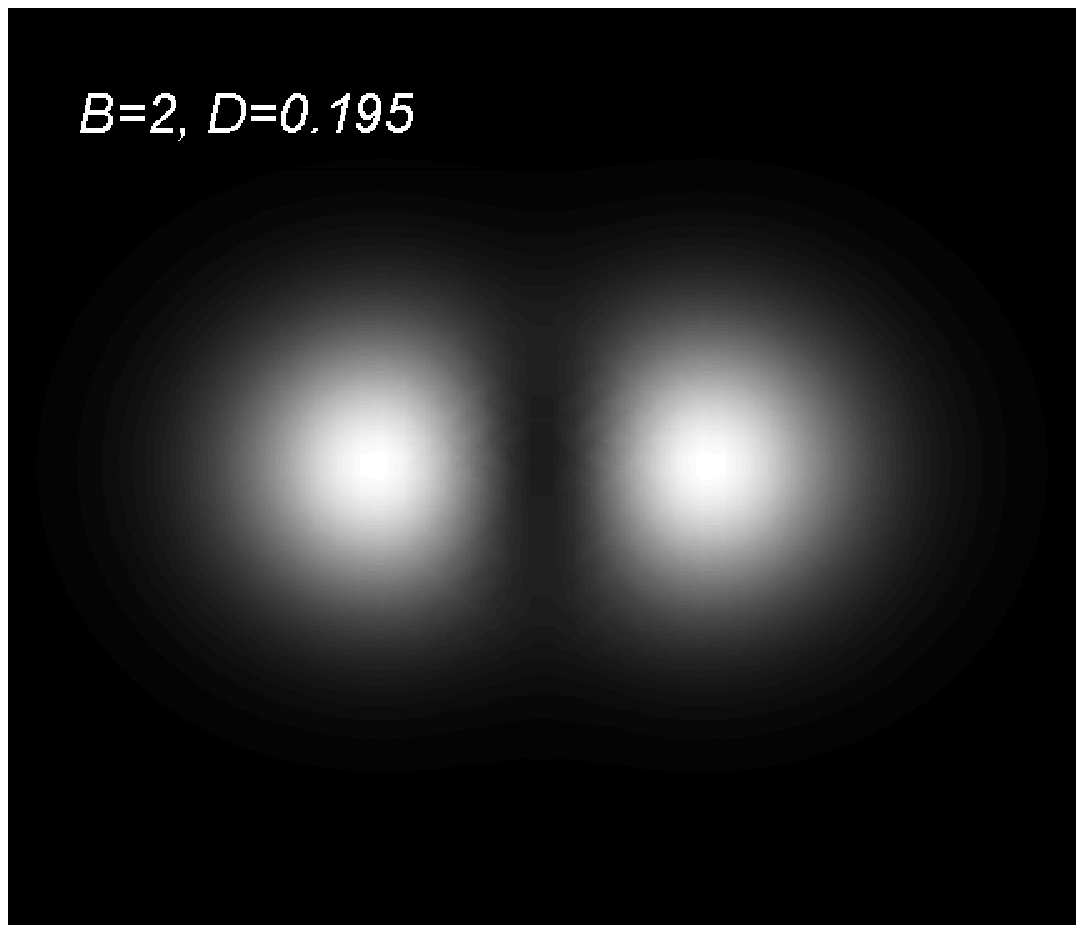} \\\hline
\end{tabular}
\caption{Energy density profiles for \(B=1\) and \(B=2\)  and for values of parameter \(D=0\)
and \(D=0.195\) (top and bottom line respectively) on the \( \rho-z\) plane. Here $\rho$  and
\(z\) span the region from -1 to 1 in units 
of ${2\sqrt{2}}/{eF_{\pi}}$.  }%
\label{Example}%
\end{figure}

Having fixed the model's parameters, we then used the algorithms of
simulated-annealing to accomplish the minimization of (\ref{EStatic}) for
values of $D$ ranging from 0 to 0.1999. As  examples of one of our
minimizations, the energy density of the $B=1,2$ Skyrmions for  values of the parameter \(D=0\) and \(D=0.195\)  are shown in
Figure \ref{Example}. For each minimization, we used a 250 by 500 point grid,
representing the standard cylindrical coordinates $z$ and $\rho$ respectively,
which provides sufficient detail for analyzing rotational and vibrational
modes. Axial symmetry is therefore implied as it is known to be a symmetry of
the $B=1 $ hedgehog and $B=2$ toroidal static solutions. These configurations  are   confirmed by the
results in Figure \ref{Example}
along with the observations that our choice of mass term  (\ref{eq-MD}) leads to non shell-like energy
densities. As one might have expected from larger mass terms,  we also see the size of the Skyrmion decreases as \(D\)  increases. Once a solution is
reached for $B=1$, we compute the matrices $M_{ij}^{0}$ and $v_{ij}$ in
equation (\ref{eq-L}) numerically. We proceed with the appropriate scaling
transformation, perform the diagonalization, and finally calculate the
eigen-energies of each vibrational mode. By the form of our diagonalization
(\ref{diag}), it is clear that we would obtain strictly three eigenvectors
depicting three types of vibration for each solution. Strikingly, but
comprehensibly, no matter the value of $D$, the three types of vibration
obtained were always of the same form.

The first type observed may be identified to the well-known breathing mode
described by the eigenvector
\begin{equation}
\frac{1}{\sqrt{3}}(1,1,1) \label{eigen1}%
\end{equation}
in Cartesian coordinates. The energy associated to this mode, $\omega_{br}$,
along with the total gives a clear picture of how the rigidity of the soliton
changes as a function of $D$. Put clearly, the lower the ratio
\begin{equation}
R_{i}=\frac{\omega_{i}}{E_{1}^{s}}, \qquad i=br,2,3, \label{ratioE}%
\end{equation}
the more the soliton is expected to be deformable, or \emph{malleable}. In
essence, the ratio (\ref{ratioE}) indicates whether or not the mass term
(\ref{eq-MD}) renders a more rigid Skyrmion, which would be desirable for the
reasons mentioned earlier. The second type, which we call $\omega_{2}$, can be
understood as a vibration along $x$ together with a simultaneous and opposite
vibration along $y$, but not $z$. Its eigenvector has the form
\begin{equation}
\frac{1}{\sqrt{2}}(1,-1,0). \label{eigen2}%
\end{equation}
The third vibrational type is characterized by a positive vibration along $x$
and $y$, together with a simultaneous opposite vibration along $z$. We call
this last type $\omega_{3}$ with eigenvector
\begin{equation}
\frac{1}{\sqrt{6}}(1,1,-2). \label{eigen3}%
\end{equation}
These eigenvectors (\ref{eigen1}), (\ref{eigen2}), and (\ref{eigen3}),
however, are idealizations of what we actually numerically obtain, even though
our numerical approach does come close to this. For example, the three
eigenvectors obtained for $D=0.8$ were (neglecting normalization)
\[
\left(
\begin{array}
[c]{c}%
4.064\\
4.064\\
4.093
\end{array}
\right)  ,\quad\left(
\begin{array}
[c]{c}%
4.210\\
-4.210\\
0
\end{array}
\right)  ,\,\text{and}\quad\left(
\begin{array}
[c]{c}%
2.446\\
2.446\\
-4.869
\end{array}
\right)  .\quad
\]
Throughout all our results, an accuracy of this type was the norm.

Table \ref{tableun} gives the energies of each vibrational mode, together with
the static and rotational energies of the $B=1$ soliton. Here the solution was
found by minimizing the static energy $E_{s}$. \begin{table}[th]
\caption{$B=1$ vibrational energies versus $D$ \emph{without} rotational
minimization (MeV)}%
\label{tableun}
\centering.%
\begin{tabular}
[c]{cccccccc}\hline\hline
$D$ & $E_{N}$ & $E^{\pi}$ & $E_{1}^{rot}$ & $E_{1}^{s}$ & $\omega_{br}$ &
$\omega_{2}$ & $\omega_{3}$\\[0.5ex]\hline
0 & 1011.53 & 26.67 & 112.12 & 899.41 & 290.08 & 648.42 & 646.63\\
0.02 & 1014.34 & 28.14 & 113.05 & 901.29 & 293.47 & 651.88 & 649.89\\
0.04 & 1017.94 & 29.85 & 114.35 & 903.59 & 297.78 & 656.55 & 654.66\\
0.06 & 1021.31 & 32.51 & 114.84 & 906.47 & 302.15 & 659.20 & 657.03\\
0.08 & 1026.87 & 35.38 & 116.64 & 910.24 & 308.67 & 665.76 & 663.10\\
0.1 & 1034.12 & 39.31 & 118.78 & 915.34 & 317.59 & 674.17 & 671.96\\
0.12 & 1044.77 & 44.65 & 122.12 & 922.65 & 330.16 & 686.46 & 684.91\\
0.14 & 1061.48 & 52.51 & 127.41 & 934.07 & 349.54 & 706.13 & 704.38\\
0.16 & 1091.90 & 65.36 & 137.33 & 954.57 & 384.15 & 742.28 & 740.63\\
0.18 & 1164.35 & 93.18 & 159.96 & 1004.39 & 463.69 & 824.65 & 823.08\\
0.19 & 1268.19 & 128.15 & 191.03 & 1077.16 & 573.29 & 937.29 & 935.27\\
0.195 & 1408.88 & 170.36 & 231.13 & 1177.75 & 716.27 & 1082.78 & 1078.47\\
0.1975 & 1591.37 & 220.0 & 280.32 & 1311.05 & 897.55 & 1265.10 & 1261.77\\
0.199 & 1907.78 & 298.78 & 361.49 & 1546.30 & 1194.68 & 1567.33 & 1546.77\\
0.1999 & 3193.16 & 578.49 & 657.96 & 2535.21 & 2311.62 & 2726.74 &
2660.59\\[1ex]\hline
\end{tabular}
\end{table}We first note that the energies of the $\omega_{2}$ and $\omega
_{3}$ modes are quite similar. Remembering that the only difference between
these is a vibration along $z$, we see that this vibration along $z$
diminishes the total energy of the vibration ever so slightly. Furthermore,
the energy of the breathing mode rises much faster as a function of $D$ than
the other two vibrational modes. Lastly, we notice that the energy of the mass
term, $E^{\pi}$, increases significantly with $D$. Yet, even with $D=0.1999$,
the mass term represents less than 23\% of the total static energy, which is
somewhat surprising if we consider the factor $(1-5D)^{-1}$ in front of the
mass term. Note that for $D=0.1999$, we still observe a typical hedgehog
configuration similar to that of Figure \ref{Example} although we are at the
frontier $(D=0.2)$ of a breakdown in our numerical procedure. Indeed, the
relatively large deviation of $\omega_{2}$ with respect to $\omega_{3}$ for
$D=0.1999$ could be an early sign of such a breakdown and caution is advised
when interpreting this last set of data.

A second set of computations were performed in parallel to the ones given
above by including (iso-)rotational energy in the energy minimization process
and therefore allowing for axial deformation of the nucleon. Such effects which have been shown to be significant  \cite{Battye:2005,Magee,Marleau2}. Thus, adding the
(iso-)rotational energy (\ref{ERot}) to the static energy (\ref{EStatic}) and
minimizing the nucleon mass instead of the static energy
\begin{equation}
E_{N}=E_{1}^{s}+E_{1}^{rot} \label{EN}%
\end{equation}
will lead to an axially deformed solution and a new set of data shown in Table
\ref{tabledeux}.
We see again that the data for \(D=0.1999\) exhibit a rather peculiar behaviour especially regarding
the value \(\omega_{br}=71.98\). Although there is no obvious signal of a numerical breakdown in the energy
and baryon number density configuration, in our opinion the data cannot be trusted. It should be noted
that the numerical breakdown is much more drastic
for higher values of \(D\) where it is generally characterized by scattered densities over the entire
grid. 
The  \(D=0.1999\) data set is nonetheless listed to illustrate the signature 
of a possible numerical breakdown.   
\begin{table}[th]
\caption{$B=1$ vibrational energies \emph{with} rotational minimization (MeV)}%
\label{tabledeux}
\centering%
\begin{tabular}
[c]{cccccccc}\hline\hline
$D$ & $E_{N}$ & $E^{\pi}$ & $E_{1}^{rot}$ & $E_{1}^{s}$ & $\omega_{br}$ &
$\omega_{2}$ & $\omega_{3}$\\[0.5ex]\hline
0 & 990.23 & 51.52 & 73.82 & 916.41 & 269.61 & 541.87 & 540.31\\
0.02 & 992.81 & 53.74 & 74.19 & 918.61 & 274.01 & 545.53 & 543.54\\
0.04 & 995.98 & 56.42 & 74.65 & 921.33 & 278.81 & 549.15 & 547.23\\
0.06 & 999.83 & 59.94 & 74.87 & 924.96 & 287.45 & 555.88 & 553.40\\
0.08 & 1004.83 & 64.24 & 75.24 & 929.60 & 297.15 & 563.21 & 560.02\\
0.1 & 1011.55 & 69.77 & 75.73 & 935.83 & 309.24 & 572.19 & 568.21\\
0.12 & 1020.92 & 77.35 & 76.04 & 944.88 & 329.01 & 586.01 & 579.82\\
0.14 & 1035.27 & 88.27 & 76.28 & 958.99 & 356.36 & 603.49 & 595.97\\
0.16 & 1060.49 & 105.10 & 77.66 & 982.82 & 395.82 & 629.89 & 616.99\\
0.18 & 1118.28 & 139.27 & 77.67 & 1040.61 & 493.25 & 680.96 & 664.99\\
0.19 & 1199.51 & 178.84 & 79.68 & 1119.84 & 603.37 & 736.22 & 718.97\\
0.195 & 1309.04 & 224.52 & 83.31 & 1225.72 & 729.46 & 803.61 & 783.27\\
0.1975 & 1452.83 & 277.16 & 89.70 & 1363.13 & 808.13 & 904.12 & 895.73\\
0.199 & 1707.07 & 360.53 & 103.82 & 1603.25 & 1003.54 & 1135.50 & 1043.64\\
0.1999 & 2309.34 & 578.49 & 151.63 & 2157.71 & 71.98 & 2102.43 &
1620.67\\[1ex]\hline
\end{tabular}
\end{table}

\begin{figure}[th]
\centering
\includegraphics[width=0.4\textwidth]{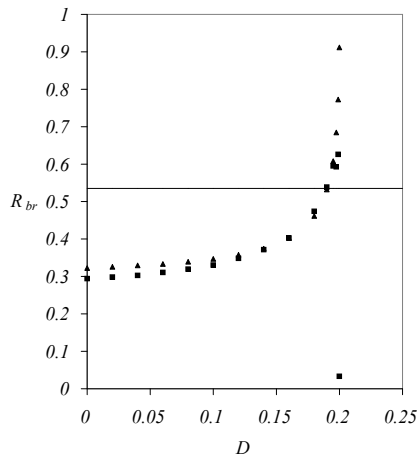}
\includegraphics[width=0.4\textwidth]{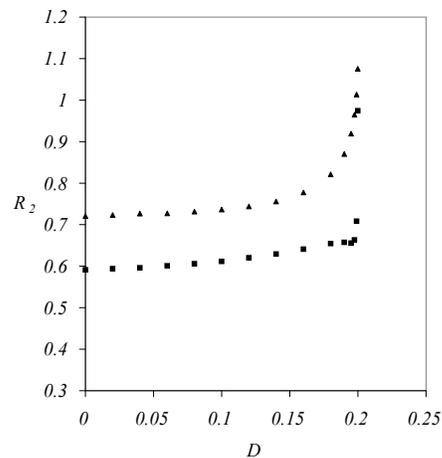}
\par
a) \hspace{2.5in} b)
\par
\quad
\par
\caption{Ratios of vibrational with respect to the static energies
$R_{i}=\frac{\omega_{i}}{E_{1}^{s}}$ as a function of the mass parameter $D$
where in a) $i $ labels the vibrational mode $i=br$ and in b) $i=2 $.
Triangles (squares) correspond to the solutions \emph{without} (\emph{with})
rotational energy in the energy minimization. The vertical solid line in a) is
the experimental value of ratio $\frac{E_{Roeper}}{E_{N}}$, where $E_{Roeper}$
is the energy of the Roeper resonance, sometimes identified with the breathing
mode of the proton, but shown here for comparison purposes only.}%
\label{figuredeux}%
\end{figure}

Notice how little the rotational energy changes with $D$ and how similarly the
vibrational energies behave as a function of $D$ with those obtained
\emph{without} rotational term in the  minimization. However, there are significant
differences between the two sets of data which are best exhibited in a plot of
the relevant ratios for both types of minimization. This is shown in Figure
\ref{figuredeux}.

\clearpage The differences between the ratios $\frac{\omega_{i}%
}{E_{1}^{s}}$ of the minimizations \emph{with} and \emph{without} rotational energy are
now clear. For the breathing mode, the curves overlap quite well. However, for
the other two types of vibration, $\omega_{2}$ and $\omega_{3}$, the added
rotational energy to the Lagrangian lowers significantly their energies.
Perhaps because $E_{1}^{rot}$ renders a slightly less oblate Skyrmion, the
vibrational energies along $\rho$ diminish accordingly. The general tendencies
of all curves however indicate that the mass term does add a certain
``rigidity'' to the field configurations as $D$ increases. What remains to be
seen is if there is a correlation between this defined rigidity and the
binding energies of $B=2$ solutions. This is done in the next section.

But before analyzing binding energies, we must mention here the complementary
results of  Lin and  Piette \cite{LP}. The authors consider the
vibrational modes of the Skyrme model with the mass term falling into the
class of eqs. (\ref{MassG}) and (\ref{MassG2}) namely of the form $Tr(U^{p}-1)
$ which differs from (\ref{eq-MD}). They follow  the time dependent approach
introduced by Barnes et al. \cite{Barnes1} to identify the vibrational modes
which assumes local instead of global scale transformations that we chose to perform our calculation. Also they use of the
rational map ansatz which seems to prevent a precise determination of the
vibrational energies.  For anyone of these reasons, their results can not be compared
directly with ours. Nonetheless they observe a clear increase in the
vibrational energies with respect to the mass term which is in qualitative
accord with our results.

\section{Binding Energies}

\begin{figure}[ptbh]
\centering
\includegraphics[width=0.4\textwidth]{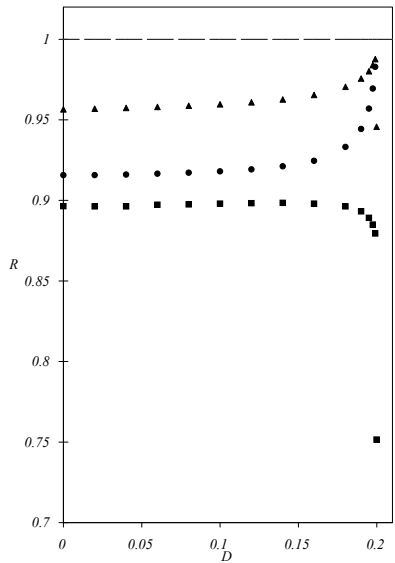}\caption{Ratios $\frac{E_{2}^{s}%
}{2E_{1}^{s}}$ (triangles) and $\frac{E_{D}}{2E_{N}}$ (squares) as a function
of the mass parameter $D$. The dashed line at $R=1$ corresponds to the limit
of instability of the bound state whereas the experimental value for
$\frac{E_{D}}{2E_{N}}=0.99979$, i.e. very close to 1.}%
\label{figuretrois}%
\end{figure}

The results for the static energies of the $B=2$ solitons are presented  in Table
\ref{tabletrois} together with their
corresponding binding energies computed from the total static energies of the $B=1$ solitons of Tables \ref{tableun}.  \begin{table}[th]
\caption{$B=2$ energies without rotational minimization (MeV) }
\centering%
\begin{tabular}
[c]{|c|ccc|ccc|}\hline\hline
$D$ & \quad & $E_{2}^{s}$ & $2E_{1}^{s}-E_{2}^{s}$ & \quad & $E_{D}$ &
$2E_{N}-E_{D}$\\\hline
0 &  & 1720.55 & 78.27 &  & 1813.51 & 209.55\\
0.02 &  & 1724.91 & 77.66 &  & 1818.19 & 210.49\\
0.04 &  & 1730.18 & 77.01 &  & 1824.65 & 211.23\\
0.06 &  & 1736.7 8 & 76.15 &  & 1832.75 & 209.86\\
0.08 &  & 1745.33 & 75.14 &  & 1843.18 & 210.57\\
0.1 &  & 1756.80 & 73.88 &  & 1857.23 & 211.00\\
0.12 &  & 1773.14 & 72.15 &  & 1876.94 & 212.60\\
0.14 &  & 1798.38 & 69.75 &  & 1907.32 & 215.63\\
0.16 &  & 1843.13 & 66.02 &  & 1961.01 & 222.79\\
0.18 &  & 1949.56 & 59.22 &  & 2087.19 & 241.51\\
0.19 &  & 2101.77 & 52.56 &  & 2265.44 & 270.94\\
0.195 &  & 2308.81 & 46.69 &  & 2505.45 & 312.32\\
$0.1975$ &  & $2580.04$ & $42.06$ &  & 2816.70 & 366.04\\
0.199 &  & 3054.49 & 38.10 &  & 3355.59 & 459.97\\
$0.1999$ &  & $4794.96$ & $275.46$ &  & 4799.64 & 1586.68\\\hline
\end{tabular}
\label{tabletrois}%
\end{table}Also shown are the results for the physical states of deuteron
$E_{D}$ versus that of the nucleon $E_{N}$. All results in Table
\ref{tabletrois} were obtained by minimizing static energies alone, i.e. 
minimization \emph{without} rotational energy,
to avoid non-axial solutions. We  recall here that we have  fixed the
$F_{\pi}$, $e$ and $m_{\pi}$ parameters of the Skyrme model leading to results
for $E_{D}$ and $E_{N}$ that are much higher than their experimental values.
While an appropriate fit of the parameters could fix this problem it is not
necessary in the context of this work since we are more interested in the
relative weight of the vibrational and binding energies than in their actual values. With
the data provided by Table \ref{tabletrois}, we define $R$, the ratio of the
energy of the $B=2$ soliton with respect to that of two isolated $B=1$
solitons. $R$ is plotted in Fig. \ref{figuretrois} for three sets of points,
\begin{equation}
R=\frac{E_{2}^{s}}{2E_{1}^{s}}%
\end{equation}
corresponding to (iso-)spinless solitons and
\begin{equation}
R=\frac{E_{D}}{2E_{N}}%
\end{equation}
for comparison of the deuteron versus two nucleons. These quantities indicates how the relative binding energies change with increasing $D$. The
dashed line represents the experimentally measured mass of the deuteron over
twice the mass of an individual nucleon (\(R=0.9998\)). 

Clearly, as $D$ approaches its
critical value of 0.2, we observe that the binding energies for \(I=J=0\)
solitons (triangles in Fig. \ref{figuretrois}) diminish considerably without ever crossing the dashed line or 
the value \(R=1\) which corresponds to the limit of instability of the $B=2$
solution.  The sharp increase in \(R \) suggest that it may be possible to  adjust the value
of $D$ such that the relative importance of the binding energy would be arbitrarily small  --- of course, the solution would still be of toroidal
form which is presumably not that of the deuteron. Some cautionary comments  are in order here. First
the last point for this set of data  shows a sharp increase of the binding energy at \(D=0.1999\). It is at
the boundary of the region  of breakdown of our numerical technique and attempts
 to
obtain stable solutions     as we push $D$ closer to 0.2 were not successful.
Also the limit \(D\rightarrow0.2\) is ill-defined and approaching this limit is physically questionable as the contributions
of the mass term which breaks the  \(SU(2)  \) symmetry gets relatively large. 

Surprisingly, the situation is more obscure for the deuteron-nucleon
data. A first set of data, based on static energy minimization (squares in Fig. \ref{figuretrois}), shows that  the relative importance of the binding energy is
quite insensitive to $D$ up to $0.17$ but increases (i.e. \(R  \) decreases) dramatically as $D$
approaches the critical value of $0.2$ contrarily to what is observed for the spinless Skyrmion. This is entirely due to the distinct
behavior of the $B=2$ and $B=1$ rotational energies. Of course rotational deformations  may affect this
behaviour. To provide some measure of that effect we present a third set of points on Fig. \ref{figuretrois}
(circles)  where \(R \) is computed with the \(B=2\) static and \(B=1\) rotationally deformed solutions. Note that  an exact full 3D  \(B=2\) rotationally deformed computation  would lower \(R,  \) and so this third set of
data represent the absolute maximum of \(R \) for the deuteron-nucleon system. Again the mass term
(\ref{eq-MD}) that we analyzed always lead to bound states ($R<1$) which is an interesting result by
itself. However
the observation of this opposite behavior with respect to the parameter $D$ for
these last two sets of data emphasizes the importance of the rotational deformations. It may well be that a full 3D
computation of  \(B=2
\) deformed solution leads to a behaviour closer to that of the second set of data (squares) but  our results does not allow to  infer on the exact  nature of the bound
state for the deuteron. Perhaps a full 3D computation allowing for non-axial deformation of the deuteron would conclude otherwise.
It is also clear that the bound states of  spinless S\textsf{a}kyrmions and deuteron-nucleon system may show completely
different behaviour.

Additionally, these results put into question the conclusions obtained by Piette and  Zakrzewski \cite{PZ}. Their results show that for a $D$
larger than approximately 0.12, the toroidal configuration is an unstable one
because its total energy is larger than twice that of a single nucleon. More
precisely, their results indicate that the ratio $\frac{E_{2}^{s}}{2E_{1}^{s}%
}$ crosses 1 at $D\approx0.12$ and continues to increase afterwards, in
contradiction with our results (Fig. \ref{figuretrois}). This discrepancy
suggests that the use of the rational map ansatz does not adequately minimize
the energies of the $B=2$ configurations, and that we cannot have too much
confidence in its ability to truly find the minimal solutions of any $B\geq2$.
Hence, the more exact simulated-annealing approach for finding solutions may
be the best way to decipher which model types are more suitable than others. It
may also be the only method for finding out why the $B=2$ Skyrmion has a
toroidal shape in the first place. There is also certainly a link to make
between the increasing rigidity  (characterized by $\frac{\omega_{br}}%
{E_{1}^{s}}$) and the changing binding energy of the solitons as a function of
$D$ which also depends on the nature of the solitons, (iso-)spinless or
deuteron-nucleons system. In our view, both characteristics are needed for a
better model, and hence adds to the appeal of  further dynamical or mass terms
such as (\ref{eq-MD}).

\section{Conclusion}

From the behaviors of the binding energies (Fig. \ref{figuretrois}) and the
breathing mode energies (Fig. \ref{figuredeux}a), we can assert that the mass
term (\ref{eq-MD}) with a non-zero value of $D$ does not quantitatively  succeed in providing
a consistent model of nuclear matter in the sense that the 
experimental value for binding energy of the deuteron was not attainable for  the values of $D $ considered
here. It also emphasize the need for rotationally deformed \(B=2 \) computations to determine the exact
behaviour as \(R \) seems to be very sensitive to rotational deformation near  $D=0.2$.
Meanwhile  it may be premature to conjecture the exact    deuteron-nucleon
 behaviour on the basis of the results  for spinless solitons. Our results have also shown that in
contrast to the rational map solutions of Piette and  Zakrzewski, the
behavior of the ratio $\frac{E_{2}^{s}}{2E_{1}^{s}}$ never crosses 1 and seems
only to approach 1 in the limit of $D\longrightarrow0.2$. Hence, this
indicates not only that the $B=2$ is a bound state for all  values of $D$ considered in this work,
but it also demonstrates the clear limitations of the rational map ansatz in
providing exact quantitative insight into how the model behaves and works. In
conjunction with the behavior of the binding energies, the ratio of
$\frac{\omega_{br}}{E_{1}^{s}}$ shows that as $D$ increases the field
configurations are more resistant to being vibrationally excited, which is
indicative of more rigid solitons and perhaps a relation between rigidity and
larger binding energies for the deuteron.

All calculations relied on fixed values for the parameters of the Skyrme model
(except for $D$) mostly for comparison purposes with previous work. These
parameters are usually fitted to reproduce the experimental values of  the
mass of the nucleon and delta or other physical quantities. Indeed, the values
of the energies we have obtained were sometimes far from those of experimental
data. Therefore, some of our conclusions may no longer hold for a more
physical choice of Skyrme parameters despite the fact that they were based on
the relative importance of each quantity. On the other hand, three dimensional
simulated-annealing programs will be needed to explore baryonic numbers beyond
2, and to confirm the validity of axially symmetric solutions for $B=2$.
Regarding the prospect of obtaining 2-nucleon shape solitons for $B=2$, it would
seem that the mass term in (\ref{eq-MD}) is not sufficient and perhaps we need
to rethink what type of lagrangian would permit such a minimal configuration. In
any event, it is also possible to add as many parameters in the mass term
(\ref{eq-MD}) with higher powers of $U$ as we wish, and this alone may lead us
towards a sounder effective theory of QCD.

This work was supported by the National Science and Engineering Research
Council of Canada.

\end{document}